\documentclass[aps,prb,twocolumn,superscriptaddress,amsmath,amssymb,10pt]{revtex4}
\usepackage{graphicx}
\usepackage{epstopdf}
\usepackage{xcolor}
\usepackage[colorlinks, linkcolor=red,citecolor=blue,urlcolor=blue]{hyperref}
\usepackage[all]{hypcap} 
\begin{document}
\bibliographystyle{apsrev}


\title{Geometry controlled superconducting diode and anomalous Josephson effect triggered by the topological phase transition in curved proximitized nanowires}

\author{A.~A. Kopasov}
\affiliation{Institute for Physics of Microstructures, Russian Academy of Sciences, 603950 Nizhny Novgorod, GSP-105, Russia}
\affiliation{Lobachevsky State University of Nizhni Novgorod, 603950 Nizhni Novgorod, Russia}
\author{A.~G.~Kutlin}
\affiliation{Max Planck Institute for the Physics of Complex Systems, D-01187 Dresden, Germany}
\author{A.~S.~Mel'nikov}
\affiliation{Institute for Physics of Microstructures, Russian Academy of Sciences, 603950 Nizhny Novgorod, GSP-105, Russia}
\affiliation{Lobachevsky State University of Nizhni Novgorod, 603950 Nizhni Novgorod, Russia}
\affiliation{Sirius University of Science and Technology, 1 Olympic Ave, 354340 Sochi, Russia}

\date{\today}

\begin{abstract}
We study the key features of the Josephson transport through a curved semiconducting nanowire. Based on numerical simulations and analytical estimates within the framework of the Bogoliubov -- de Gennes equations we find the ground state phase difference
$\varphi_0$ between the superconducting leads
tuned by the spin splitting field $h$ driving the system from the topologically trivial to the nontrivial superconducting state. The phase $\varphi_0$ vanishes for rather small $h$, grows in a certain field range around the topological transition and then saturates at large $h$ in the Kitaev regime. Both the subgap and continuum quasiparticle levels are responsible for the above behavior of the anomalous Josephson phase. It is demonstrated that the crossover region on $\varphi_0(h)$ dependencies reveals itself in the superconducting diode effect. The resulting tunable phase battery can be used as a probe of topological transitions in Majorana networks and can become a useful element of various quantum computation devices.
\end{abstract}


\maketitle

\section{Introduction}\label{introduction}
Semiconducting nanowires with strong spin-orbit interaction and induced superconductivity (also known as the Majorana nanowires) have been intensively studied in the past decade as the perspective devices for the realization of the Majorana states~\cite{Kitaev, Nayak, Alicea1, Aasen, Alicea2, Elliot, Aguado}. A lot of experimental works have been carried out to observe various signatures of the Majorana states in such systems, especially the zero-bias conductance peak in tunneling transport~\cite{Das2012,Churchill2013, Finck2013, Albrecht2016, Nichele2017, Zhang2018, Grivnin2019, Bommer2019}. The need to provide an unambiguous evidence in favor of the topological superconductivity (see Ref.~\cite{Pan2020} and references therein) stimulates the exploration of different experimental setups and tests for probing the topology of the superconducting condensate. One of the promising alternatives is to study the manifestations of the nontrivial topology in Josephson junctions formed by proximitized semiconducting nanowires~\cite{Cheng2012, Pikulin2012, San-Hose2013, Cayao2015, Marra2016, Peng2016, Nesterov2016, Cayao2017, Cayao2018, Murthy2020, Kjaergaard2017,Goffman2017}. In particular, one of the signatures of nontrivial topology in such junctions is the jump of the derivative $\partial I_c/\partial H$ at the topological transition~\cite{San-Hose2013}. Here $I_c$ is the critical supercurrent and $H$ is the external magnetic field directed along the wire axis. On the other hand, possible applications of the Majorana nanowires to quantum computing demand the fabrication of networks of rather complex configurations~\cite{Alicea1}. Some experimental progress has already been made along this line. Indeed, several groups have reported their progress in fabrication of advanced nanowire devices including nanowire ``hashtags'' based on InSb/Al platform~\cite{Gazibegovic2017} and InAs wurtzite nanocrosses~\cite{Krizek2017}. Thus, the analysis of the geometry-induced transport effects in proximitized nanowire junctions is an important and timely problem.

\begin{figure}[htpb]
\centering
\includegraphics[scale = 0.5]{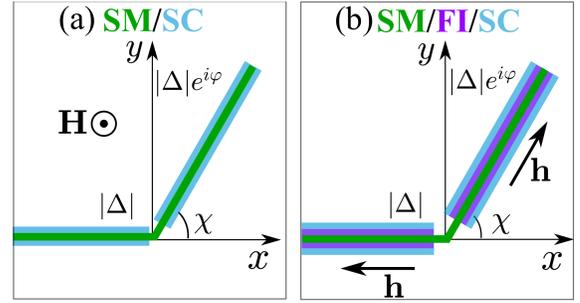}
\caption{\label{Fig:model_system} Schematic picture of curved nanowire junctions with the superconducting phase difference $\varphi$ and the geometrical offset angle $\chi$. Panels (a) and (b) show orientations of the spin splitting field considered in the present work. In panel (a) semiconductor/superconductor (SM/SC) nanowires are subjected to the external magnetic field $\mathbf{H}$ while on the panel (b) a textured spin splitting field $\mathbf{h}$ is induced in nanowires via a spin-dependent tunneling through the ferromagnetic insulator (FI).}
\end{figure}

It has been recently predicted that the Josephson junctions formed by proximitized nanowires connected with an offset angle [see Fig.~\ref{Fig:model_system}(a)] should exhibit an anomalous Josephson effect~\cite{Ying2017, Spanslatt2018, Kutlin2020}. Such behavior stems from the geometry-induced switching of the direction of the spin-orbit field which, in turn, leads to the appearance of the anomalous phase shift in the ground state of such junctions. Note that the systems exhibiting the anomalous Josephson effect are in the focus of current research work~\cite{Buzdin2008, Margaris2010, Yokoyama2014, Dolcini2015, Konschelle2015, Robinson2019, Pal2019, Szombati2016, Mayer2020, Strambini2020} due to the possibility of their use as phase batteries. The phase battery is known to be an important device of superconducting electronics which provides a constant phase shift between two superconductors in a quantum circuit~\cite{Linder2015}. In the case of curved nanowire junctions the resulting anomalous phase shift is shown to be equal to the offset angle $\chi$.
However, the analysis of this effect has been restricted to the limit of large Zeeman fields and included only the contribution from the subgap quasiparticle states which are most sensitive to the topology changes. This approach can not be used in the vicinity of the topological transition when the Kitaev model for the induced $p$-wave superconducting correlations in the low-energy spin split subband is no more valid and both spin split subbands should be taken into account~\cite{vonOppen2014}. Another crucial aspect for the analysis of the Josephson effect in topological nanowire junctions is that both subgap and continuum states contribute to supercurrent even in the case of short junctions with the sizes less than the superconducting coherence length~\cite{San-Hose2013, Cayao2015, Nesterov2016, Cayao2017, Cayao2018, Murthy2020}. In particular, it has been shown that the contribution of continuum states to supercurrent is responsible for a finite critical current at the topological transition when the excitation spectrum of proximitized nanowires is gapless~\cite{San-Hose2013}. 
The main goal of our work is to study the key features of the Josephson transport through a curved semiconducting nanowire within the full range of the Zeeman fields covering both topologically trivial and nontrivial regions of the phase diagram.

Based on the Bogoliubov -- de Gennes (BdG) equations for curved nanowire junctions, we reveal the magnetic field driven crossover from conventional to anomalous Josephson effect as the system undergoes the topological phase transition. We show that both subgap and continuum states are responsible for such behavior. The distinctive features of the Josephson effect associated with the topological superconductivity are studied for two inequivalent orientations of the spin splitting field shown in panels (a) and (b) of Fig.~\ref{Fig:model_system}. These configurations differ by the origin of the spin splitting field induced in a semiconductor. First, we consider the case when a curved nanowire junction is placed in the external magnetic field directed perpendicular to the substrate and, thus, the spin splitting field appears due to the Zeeman effect [see Fig.~\ref{Fig:model_system}(a)]. We find that in this regime the anomalous phase shift vanishes for small enough Zeeman fields and saturates at the geometrical offset angle for large fields in the topologically nontrivial phase.
We have also considered the possibility of a textured profile of the spin splitting field directed parallel and antiparallel to the nanowire axis in different parts of the system [see Fig.~\ref{Fig:model_system}(b)]. This configuration has been motivated by recent experiments~\cite{Vaitikenas2020} on hybrid ferromagnetic nanowires InAs/EuS/Al in which the spin splitting field can be induced in a semiconductor due to the spin-dependent tunneling through an additional layer of ferromagnetic insulator EuS. In such hybrid structures the resulting spin splitting field is of exchange origin. It has been recently shown that the physical mechanism responsible for the existence of the topological phase and the criterion of the topological transition in ferromagnetic hybrid nanowires are the same as in semiconductor/superconductor nanowires~\cite{Maiani2020,Langbehn2020}. Thus, both these systems can be described within the framework of commonly used phenomenological theory~\cite{LutchynOr,OregOr}. Quite opposite to the previous case, we show that in this regime the crossover region extends far into the topologically nontrivial phase, so that the anomalous phase shift doesn't saturate at the geometrical offset angle even for rather large spin splitting fields. The analysis of the behavior of the anomalous phase demonstrates that this phase can be tuned by the spin splitting field and, thus, the curved nanowire junctions may realize a tunable phase battery.

We find that for both orientations of the spin splitting field the presence of the above-mentioned crossover reveals itself in the superconducting diode effect: the magnitude of the critical current depends on the direction of the applied current. The superconducting diode effect enables directional charge transport without energy loss at low temperatures and has been recently observed in artificial superlattices Nb/Va/Ta~\cite{Ando2020}. The possibility of the superconducting diode effect has been also predicted in a variety of Josephson systems including Josephson junctions through the helical edge states of a quantum spin-Hall insulator~\cite{Dolcini2015}, ferromagnetic Josephson junctions with spin-active interfaces~\cite{Margaris2010,Pal2019}, and Josephson junctions through a straight multi-channel semiconducting nanowire with strong spin-orbit coupling in the presence of magnetic field~\cite{Yokoyama2014}. Note that according to previous theoretical works the appearance of nonreciprocal transport in junctions through straight semiconducting nanowires requires multi-channel regime. Our results show that the curved geometry allows nonreciprocal transport even through a single-channel nanowire. Moreover, it is the system geometry which controls the parameters of nonreciprocal transport.

The manuscript is organized as follows. In Sec.~\ref{basic_equations} we introduce the model and basic equations. In Sec.~\ref{section3} we present the analytical results and qualitative arguments elucidating the geometry-induced transport effects under consideration. In Sec.~\ref{results_and_discussion} the results of numerical simulations are presented and discussed. In Sec.~\ref{discussion} we discuss the possibility of experimental verification of our predictions. Finally, the results are summarized in Sec.~\ref{summary}.

\section{Basic equations}\label{basic_equations}
Our analysis is based on BdG equations~\cite{Kutlin2020}:
\begin{subequations}\label{BdG_general}
\begin{align}
 \check{H}_{\text{BdG}}(s)\Psi(s) = E\Psi(s) \ ,\\
\label{BdG_matrix}
\check{H}_{\text{BdG}}(s) = \xi(s)\check{\tau}_z + \mathbf{h}(s)\hat{\boldsymbol{\sigma}} \\
\nonumber
+ |\Delta|[\check{\tau}_x\cos\varphi(s) - \check{\tau}_y\sin\varphi(s)]- \frac{\alpha}{2}\left\{\hat{\sigma}_n(s), p\right\}\check{\tau}_z \ .
\end{align}
\end{subequations}
Here the coordinate $s$ parametrizes the wire location $\mathbf{r}(s) = [x(s),y(s)]$, $\xi(s) = p^2/2m - \mu(s)$, $p = -i\hbar\partial_s$, $\hbar$ is the reduced Planck constant, $m$ is the effective mass, $\mu(s)$ is the chemical potential profile, $\alpha$ is the spin-orbit coupling constant, $\left\{A,B\right\}$ denotes the anticommutator of $A$ and $B$, $\hat{\sigma}_n(s) = \hat{\sigma}_x\sin\chi(s)-\hat{\sigma}_y\cos\chi(s)$, $\chi(s)$ is the geometrical phase defined as the angle between the tangent vector at $\mathbf{r}(s)$ and the $x$-axis, $\hat{\sigma}_i$ ($i = x,y,z$) are the Pauli matrices acting in the spin space, $\check{\tau}_i$ ($i = x,y,z$) are the Pauli matrices acting in the electron-hole space, and $\Psi(s) = [u_{\uparrow}(s), u_{\downarrow}(s), v_{\downarrow}(s), v_{\uparrow}(s)]^{\text{T}}$ is the quasiparticle wave function. Throughout the work we use the following profiles of the spin splitting field:
\begin{subequations}
 \begin{align}
  \label{Zeeman_profile_perpendicular}
  \mathbf{h}(s) = \left[0, 0, -h_{\perp} \right]  \ ,\\
  \label{Zeeman_profile_inplane}
  \mathbf{h}(s) = h_{||}{\rm sgn}(s)\left[\cos\chi(s), \sin\chi(s), 0\right] \ ,
 \end{align}
\end{subequations} 
which are shown schematically in panels (a) and (b) of Fig.~\ref{Fig:model_system}, respectively. Without loss of generality, we take $h_{\perp},h_{||} \geq 0$. We emphasize that the presence of the wire curvature doesn't affect the criterion of the topological transition in the system. The system is in the topologically trivial (nontrivial) phase when the condition $h_{\perp},h_{||} < h_c$ ($h_{\perp},h_{||} > h_c$) is satisfied in both nanowire parts. Hereafter, $h_c = \sqrt{\mu^2 + |\Delta|^2}$. We restrict ourselves to the study of short Josephson junctions and use step-wise profiles for both superconducting and geometrical phases $\varphi(s) = \varphi\Theta(s)$, $\chi(s) = \chi\Theta(s)$ where $\Theta(x)$ is the Heaviside step function. 

BdG equations~(\ref{BdG_general}) should be supplemented by the appropriate boundary conditions at the turning point of the nanowire which take into account various mechanisms of quasiparticle scattering. In real experimental situation these mechanisms can include  spin-independent scattering~\cite{Blonder1982}, spin-dependent scattering at spin-active~\cite{Millis1988,Fogelstrom2000} and/or spin-orbit active interfaces~\cite{Sun2015}. For simplicity, here we restrict ourselves to consideration of two scattering mechanisms: (i) the spin-dependent scattering which appears due to the combined effect of spin-orbit coupling and the wire curvature and (ii) the spin-independent scattering. The former case has been addressed numerically (the details of our numerical simulations are presented in Appendix~\ref{details_numerical_simulations}). The effects of the spin-independent scattering have been analyzed within our analytical approach developed for the first system configuration shown in Fig.~\ref{Fig:model_system}(a). Note that we completely neglect possible spin-dependent scattering effects associated, e.g., with spin-active interfaces.

The zero-temperature current-phase relation $I_s(\varphi)$ has been obtained from the excitation spectrum of the system~\cite{Beenakker}
\begin{equation}\label{supercurrent_equation}
 I_s(\varphi) = \frac{2e}{\hbar}\frac{\partial E_{tot}}{\partial \varphi} \ , \ \ \ E_{tot}(\varphi) = -\frac{1}{2}\sum_{n}E_n(\varphi) \ ,
\end{equation}
where $e$ is the electron charge, $E_{tot}$ is the junction energy at zero temperature and the summation should be carried out over all the positive eigenvalues $E_n$ of the problem~(\ref{BdG_general}). Let us denote the superconducting phase difference corresponding to the ground-state energy (i.e. the minimum of the function $E_{tot}(\varphi)$) as the anomalous phase $\varphi_0$. It is easy to see from Eqs.~(\ref{supercurrent_equation}) that in the case when $E_{tot}(\varphi)$ dependence has a single minimum, the anomalous phase shift can be also extracted from $I_s(\varphi)$ dependencies using the relations:
\begin{equation}\label{anomalous_phase_conditions}
        I_s(\varphi_0) = 0 \ , \ \ \ \frac{\partial I_s}{\partial \varphi}\biggl|_{\varphi=\varphi_0} > 0 \ .
\end{equation}

Before proceed further, let us briefly comment on the validity of our approach which is similar to the one used previously in Refs.~\cite{Cayao2015,Cayao2017,Cayao2018} for the junctions with straight nanowires ($\chi=0$). First, our analysis is based on a single-channel model for proximitized nanowires~\cite{LutchynOr,OregOr} and, thus, here we don't address the questions related to multiband occupancy in the semiconducting core. Second, in the present work we use a phenomenological model of the proximity effect which is valid in the case of small tunneling rate $\Gamma \ll \Delta_0$ where $\Gamma \propto t^2\nu_0$, $t$ is the tunneling matrix element, $\nu_0$ is the normal-state density of states at the Fermi level in the metal shell, and $\Delta_0$ is the energy gap in the parent superconductor~\cite{Kopnin2011,Stanescu2011}. Within this limit, the modulus of the gap parameter is equal to the tunneling rate $|\Delta|=\Gamma$. The correct description of the proximity effect also requires to take into account the low-energy renormalization of the quasiparticle spectrum due to the diagonal matrix elements of the tunneling self-energy in the electron-hole space~\cite{Kopnin2011,Stanescu2011}. Assuming that the energies of relevant excitations contributing to supercurrent are much less than the energy gap $\Delta_0$ in the parent superconductor one can neglect the spectrum renormalization determined by the small parameter $\Gamma/\Delta_0$.

\section{Quasiparticle spectra and supercurrent. Qualitative consideration}\label{section3}
\subsection{Perpendicular Zeeman field}
Here we discuss some approximations and analytical results for the system configuration shown in Fig.~\ref{Fig:model_system}(a). We note that a lot of insight into the physics of the anomalous Josephson effect in curved junctions can be gained from the analysis of the energy spectrum of a homogeneous proximitized nanowire. For this purpose, it is convenient to rewrite the BdG model~(\ref{BdG_general}) in the basis of the so-called helical states~\cite{Alicea2,San-Hose2013}. In the helical basis, the BdG equations for the spinful proximitized nanowire are reduced to the model describing two one-dimensional superconductors with the intraband $p$-wave and the interband $s$-wave pairing (see also Appendix~\ref{appendix_B}). The resulting equations allow a simple analytical treatment in the limit $\mu \gg |\Delta|, m\alpha^2$ and $h_{\perp}\gtrsim |\Delta|$. Within this limit, it is sufficient to take into account only the $p$-wave pairing in each spin split subband while the interband $s$-wave pairing can be neglected~\cite{San-Hose2013}. The effect of the latter pairing becomes small due to a small coupling between the quasiparticle wave functions in different subbands characterized by rather strong mismatch of the Fermi momenta. As a result, the problem is reduced to two independent $p$-wave superconductors described by the following BdG models (see Appendix~\ref{appendix_B} for details):
\begin{subequations}\label{simplified_equations}
 \begin{align}
 \label{BdG_subbands_simplified}
 \check{H}_{\eta}(s) = \begin{bmatrix}\xi(s) + \eta h_{\perp}&\Delta_{p\eta}(s)\\ \Delta_{p\eta}^{\dagger}(s)&-\xi(s) -\eta h_{\perp}\end{bmatrix} \ ,\\
 \label{pairing_terms_simplified}
 \Delta_{p\eta}(s) = \frac{i\alpha}{2h_{\perp}}\left\{\Delta(s)e^{i\eta \chi(s)}, p\right\} \ ,
 \end{align}
\end{subequations}
where $\eta = \pm$ is the subband index. Thus, the total supercurrent through a curved nanowire can be approximated by a sum of supercurrents carried by the Andreev bound states produced by each subband.

\begin{figure}[htpb]
\centering
\includegraphics[scale = 0.95]{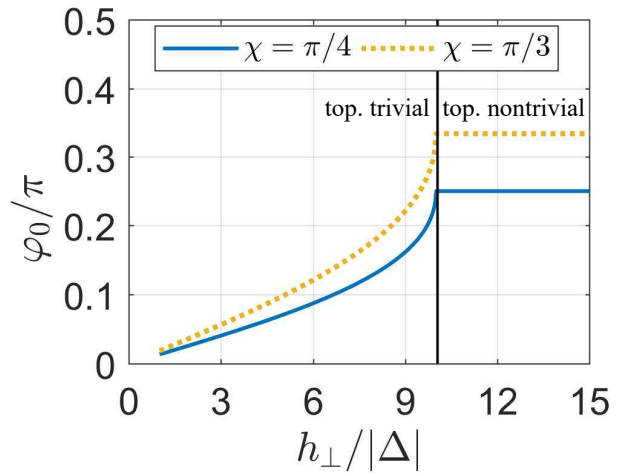}
\caption{\label{Fig:theory} Typical analytical $\varphi_0(h_{\perp})$ dependencies given by Eqs.~(\ref{supercurrent_equation}) and~(\ref{subgap_spectrum_simplified}). We take $\mu = 10|\Delta|$ and $m\alpha^2 = 0.2|\Delta|$ for both nanowires. The system undergoes the topological phase transition at $h_{\perp} = \sqrt{\mu^2 + |\Delta|^2}$ (denoted as a black solid line).}
\end{figure}

Particularly simple expressions for the energy spectra of the subgap quasiparticle states and the current-phase relation can be obtained for transparent junctions by treating Eqs.~(\ref{simplified_equations}) within the quasiclassical approximation. We get the following results for the subgap energy spectrum (see Appendix~\ref{equations_derivation} for the derivation):
\begin{subequations}\label{subgap_spectrum_simplified}
\begin{align}  
  \label{subgap_spectrum}
  E_{\eta}(\varphi) = \pm |\Delta_{\eta}|\cos\left[\left(\varphi + \eta\chi\right)/2\right] \ ,\\
  \label{simplified_gaps_vs_Zeeman}
  |\Delta_{\eta}| = |\Delta|{\rm Re}\left[\sqrt{2m\alpha^2(\mu-\eta h_{\perp})}\right]/h_{\perp} \ .
\end{align}
\end{subequations}
Here $|\Delta_{\eta}|$ is the induced superconducting gap in the quasiparticle spectrum of the subband $\eta$. Substituting Eq.~(\ref{subgap_spectrum_simplified}) into Eq.~(\ref{supercurrent_equation}) and performing the summation over the subband index, we derive the current-phase relation at zero temperature:
\begin{eqnarray}\label{supercurrent_analytical}
 I_s(\varphi) = \\
 \nonumber
 \sum_{\eta = \pm} \mathcal{I}_{\eta}\sin\left[(\varphi+\eta\chi)/2\right]{\rm sgn}\left\{\cos\left[(\varphi+\eta\chi)/2\right]\right\} \ ,  
\end{eqnarray}
where $\mathcal{I}_{\eta} = e|\Delta_{\eta}|/2\hbar$ is the subband critical current. One can clearly see from Eq.~(\ref{subgap_spectrum}) that the dispersion of the subgap levels related to different subbands have the opposite phase shifts ($\pm \chi$). Another crucial aspect for subsequent analysis is that the gaps $|\Delta_{\eta}|$ demonstrate qualitatively different behavior with respect to the Zeeman field due to the mismatch of the Fermi momenta in spin split subbands $p_{F\eta} = {\rm Re}[\sqrt{2m(\mu - \eta h_{\perp})}]$. Indeed, Eq.~(\ref{simplified_gaps_vs_Zeeman}) suggests that the gap $|\Delta_-|$ is finite within the full range of considered Zeeman fields. However, the gap $|\Delta_+|$ decreases upon the increase in the Zeeman splitting from zero and then vanishes at the topological transition (note that in the limit $\mu \gg |\Delta|$ the topological transition occurs at $h_{\perp} \approx \mu$). For $h_{\perp}>\mu$ the bottom of the normal-state $\eta = +$ subband is located above the Fermi level and, thus, this subband should be removed from the low-energy problem in the topologically nontrivial phase. Calculations of the junction energy using the subgap spectra~(\ref{subgap_spectrum}) reveal the presence of two local minima of the total energy vs. the superconducting phase difference and quite complex behavior of the anomalous phase shift. In particular, the anomalous phase $\varphi_0$ vanishes in the limit $h_{\perp}\to 0$, grows in a certain field range around the topological transition and saturates at the geometrical offset angle $\varphi_0(h_{\perp}) = \chi$ for $h_{\perp}\geq \mu$ (see Fig.~\ref{Fig:theory}). We emphasize that except the vicinity of the topological phase transition Eq.~(\ref{supercurrent_analytical}) gives us the anomalous phase value which is in a good agreement with the results of our numerical simulations for $\mu \gg |\Delta|$ (see Section~\ref{results_and_discussion}).  

We continue with the study of key transport characteristics of the Josephson effect. In particular, we investigate the superconducting diode effect and following Ref.~\cite{Yokoyama2014} we introduce $I_{c+}$ (maximum of the Josephson current) and $I_{c-}$ (absolute value of the minimum current). Note that the possiblity of nonreciprocal transport through a curved proximitized nanowire is not forbidden by the symmetry considerations due to the fact that the inhomogeneous Rashba term in Eq.~(\ref{BdG_matrix}) breaks the inversion symmetry $s\to -s$. Thus, for a finite spin-orbit coupling constant $\alpha$, the system aquires a preferential direction which is exactly the direction of the current flow.  
Considering the current-phase relation~(\ref{supercurrent_analytical}), we derive the following expressions for the critical currents:
\begin{subequations}\label{critical_currents_analytical}
\begin{align}
 I_{c+} = \begin{cases}\mathcal{I}_-\cos\chi + \mathcal{I}_+ \ , \ \ \ 0\leq \chi \leq \chi_0 \ ,\\
            \mathcal{I}_- - \mathcal{I}_+\cos\chi \ , \ \ \ \chi_0 < \chi \leq \pi \ ,
           \end{cases}\\
 I_{c-} = \begin{cases}|\mathcal{I}_- + \mathcal{I}_+\cos\chi| \ , \ \ \ 0\leq \chi \leq \pi - \chi_0 \ ,\\
            |\mathcal{I}_-\cos\chi - \mathcal{I}_+| \ , \ \ \ \pi - \chi_0 < \chi \leq \pi \ .
           \end{cases}    
\end{align}
\end{subequations}
Here $\chi_0 = 2\arctan(\sqrt{\mathcal{I}_+/\mathcal{I}_-})$.

Let us now discuss the effect of an additional spin-idependent scattering on the junction characteristics. For this purpose, we introduce a potential barrier at the turning point of the nanowire which is described by the term $V_0\check{\tau}_z\delta(s)$ added into the Hamiltonian~(\ref{BdG_subbands_simplified}). We find that the spin-independent scattering doesn't affect the functional form of the current-phase relation shown in Eq.~(\ref{supercurrent_analytical}) and leads to the following renormalization of the subband critical currents $\mathcal{I}_{\pm} = e|\Delta_{\eta}|\sqrt{D_{\eta}}/\hbar$. Here $D_{\eta} = (1 + Z^2_{\eta})^{-1}$ is the transparency of each $p$-wave Josephson junction and $Z_{\eta} = mV_0/\hbar p_{F\eta}$ is the corresponding barrier strength parameter. Thus, all our qualitative predictions regarding the junction transport characteristics remain valid in this case, though, the increase in the barrier strength suppresses the supercurrent. As for the spin-dependent scattering, one can expect that it can affect the above contributions to the Josephson current arising from different subbands. Still, we expect our predictions can survive even in this situation except for a very exotic case when this scattering completely suppresses the current contribution from one of the spin channels.

The above-described generalization of our results in the presence of the spin-independent scattering can be used to clarify the relation between the features of the Josephson transport in curved nanowire junctions and the Majorana physics. Indeed, one can easily get the following expression for the subgap spectrum in the regime $h_{\perp} \gtrsim \sqrt{\mu^2 + |\Delta|^2}$ 
\begin{equation}
E_-(\varphi) = \pm |\Delta_-| \sqrt{D_-}\cos[(\varphi - \chi)/2] \ .
\end{equation}
Note that for $\chi = 0$ the above expression coincides with the well-known expression describing the energy spectra of the quasiparticle levels in topologically nontrivial Josephson junctions~\cite{Kitaev, Kwon2004, Tarasinski2015}. One can clearly see that in the case of zero transparency we get a pair of isolated Majorana modes localized at the turning point of the nanowire. The increase in the junction transparency results in the hybridization of these low-energy quasiparticle states. The above equation also demonstrates that such hybridization is controlled by the junction geometry through the dependence on the geometrical offset angle $\chi$. We emphasize that the contribution of the above-described subgap state to supercurrent is one of two contributions responsible for the appearance of geometry-induced transport effects under consideration. Thus, the Josephson transport phenomena determined by the above Majorana pair localized at the turning point definitely reflect a general topology change.

\subsection{Textured spin splitting field}
Before proceed further, let us note that the system doesn't exhibit the anomalous Josephson effect when the direction of the spin splitting field follows the direction of the nanowire axis
\begin{equation}\label{Zeeman_inplane_aligned}
 \mathbf{h}(s) = h_{||}\left[\cos\chi(s), \sin\chi(s), 0\right] \ .
\end{equation}
It is straightforward to show that for the above profile of the spin splitting field the BdG operators $\check{H}_{{\rm BdG}}(s,\varphi)$ and $\check{H}_{{\rm BdG}}(s,-\varphi)$ are related by a unitary transformation and, thus, the excitation energies are even functions of the superconducting phase difference
\begin{subequations}
\begin{align}
 \check{H}_{{\rm BdG}}(s,-\varphi) = \hat{\sigma}_x\mathcal{P}\check{H}_{{\rm BdG}}(s,\varphi)\mathcal{P}\hat{\sigma}_x \ ,\\
 \label{spectrum_even_function}
 E_n(\varphi) = E_n(-\varphi) \ .
 \end{align}
\end{subequations}
Here $\mathcal{P}$ denotes the parity inversion operator $\mathcal{P}f(s) = f(-s)$. Combining Eqs.~(\ref{spectrum_even_function}) and~(\ref{supercurrent_equation}), we get that the supercurrent satisfies the relation $I_s(\varphi) = -I_s(-\varphi)$ and the anomalous Josephson current is equal to zero due to the symmetry constraint. Thus, in order to get a finite anomalous phase shift it is necessary to consider the change in the direction of the spin splitting field relative to the nanowire axis in different parts of the system.

In the rest of this subsection we consider the case when the direction of the spin splitting field is parallel and antiparallel to the nanowire axis in different parts of the system [see Fig.~\ref{Fig:model_system}(b)].
Note that the main difference between two profiles of the spin splitting field considered in our work is that the textured spin splitting field shown in Fig.~\ref{Fig:model_system}(b) results in different spin structures for the quasiparticle wave function in straight nanowire parts. Thus, a simplified theoretical description which has been implemented for the configuration shown in Fig.~\ref{Fig:model_system}(a) is no longer applicable since the boundary conditions at the junction couple the eigenstates from both spin split subbands. This fact complicates further analytical progress for this type of system configuration and forces us to rely only on qualitative arguments derived from the results of numerical simulations.
The complex behavior of the anomalous phase shift and the superconducting diode effect appear due to a nontrivial form of the current-phase relation which consists of two contributions: (i) the contribution of the subgap quasiparticle states and (ii) the contribution of the continuum levels (high-energy resonant states). 
Regarding the subgap quasiparticle states, we note that for large spin splitting fields $h_{||} \gtrsim \mu$ this contribution strongly depends on the geometrical offset angle $\chi$ owing to spin-filtering properties of a curved junction with a magnetic texture. As for the resonant states,  our further numerical simulations reveal that this contribution comes primarily from the energy range $E \in [h_{||} - |\Delta|, h_{||} + |\Delta|]$.

\section{Numerical simulations}\label{results_and_discussion}

\subsection{Perpendicular Zeeman field}

\begin{figure}[htpb]
\centering
\includegraphics[scale = 0.65]{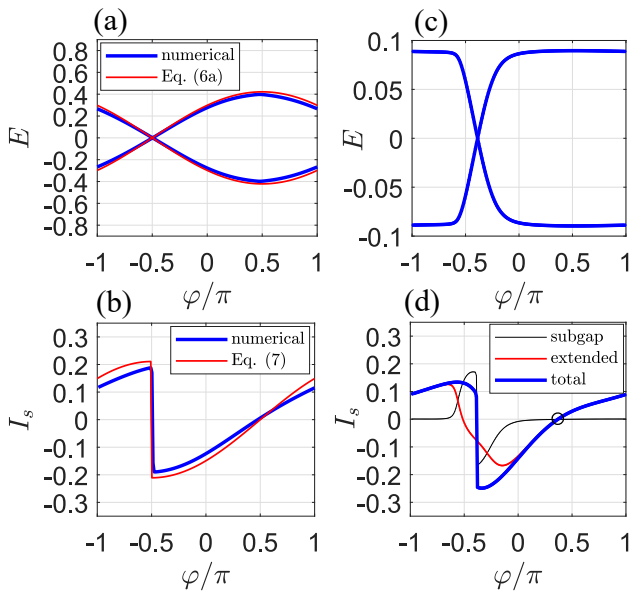}
\caption{\label{Fig:perpendicular_sub_gap_states_supercurrent} Typical subgap spectra and current-phase relations calculated for the Zeeman field profile~(\ref{Zeeman_profile_perpendicular}). Hereafter, we use $|\Delta|$ as the energy unit and $e|\Delta|/\hbar$ as the current unit. We take $\chi = \pi/2$, $\mu = |\Delta|$ and $m\alpha^2 = 0.2|\Delta|$ for both nanowires. The system undergoes the topological phase transition at $h_{\perp} = \sqrt{2}|\Delta|$. Panels (a) and (b) correspond to $h_{\perp} = 3|\Delta|$ while panels (c) and (d) correspond to $h_{\perp} = 1.5|\Delta|$. Black circles denote the superconducting phase difference at which the energy $E_{tot}$ of the junction reaches its minimal value (the anomalous phase shift $\varphi_0$).}
\end{figure}

We proceed with the discussion of the results of numerical simulations. Here we consider the model profile of the Zeeman field given by Eq.~(\ref{Zeeman_profile_perpendicular}). Typical subgap spectra and current-phase relations are shown in Fig.~\ref{Fig:perpendicular_sub_gap_states_supercurrent}. First, we address the case of large Zeeman fields $h_{\perp}\gtrsim h_c$ [panels (a) and (b) in Fig.~\ref{Fig:perpendicular_sub_gap_states_supercurrent}]. Within our numerical approach we find that in this regime the contribution of the subgap states to supercurrent well exceeds the contribution of the continuum states and, thus, the analysis of the subgap quasiparticle spectra is enough to fully describe the features of the dc Josephson effect. We perform a direct comparison of the results of our numerical simulations with analytical expressions for the subgap spectrum and the current-phase relation given by Eqs.~(\ref{subgap_spectrum_simplified}) and~(\ref{supercurrent_analytical}), respectively. The plots shown in panels (a) and (b) of Fig.~\ref{Fig:perpendicular_sub_gap_states_supercurrent} demonstrate a good agreement between analytical and numerical results.

Approaching the topological phase transition from the topologically nontrivial region of the phase diagram, we find that the analytical approach used in Refs.~\cite{Spanslatt2018,Kutlin2020} is no more valid and one has to take into account the contributions to the supercurrent from both the subgap and the continuum states. Typical subgap spectrum and the current-phase relation near the topological transition are shown in panels (c) and (d) of Fig.~\ref{Fig:perpendicular_sub_gap_states_supercurrent}. In particular, Fig.~\ref{Fig:perpendicular_sub_gap_states_supercurrent}(c) demonstrates that the dispersion of the subgap levels with respect to $\varphi$ is nonvanishing only in the vicinity of the zero-energy crossing. As a result, the contribution of the subgap states to the supercurrent near the topological phase transition is negligibly small within a wide range of $\varphi$ and a finite anomalous phase shift appears exactly due to the contribution of the continuum states [see Fig.~\ref{Fig:perpendicular_sub_gap_states_supercurrent}(d)]. Quite opposite to large fields $h_{\perp}$ well above the field $h_c$, the results in Fig.~\ref{Fig:perpendicular_sub_gap_states_supercurrent}(d) indicate that the anomalous phase shift deviates from the geometrical offset angle near the topological phase transition.

\begin{figure}[htpb]
\centering
\includegraphics[scale = 0.9]{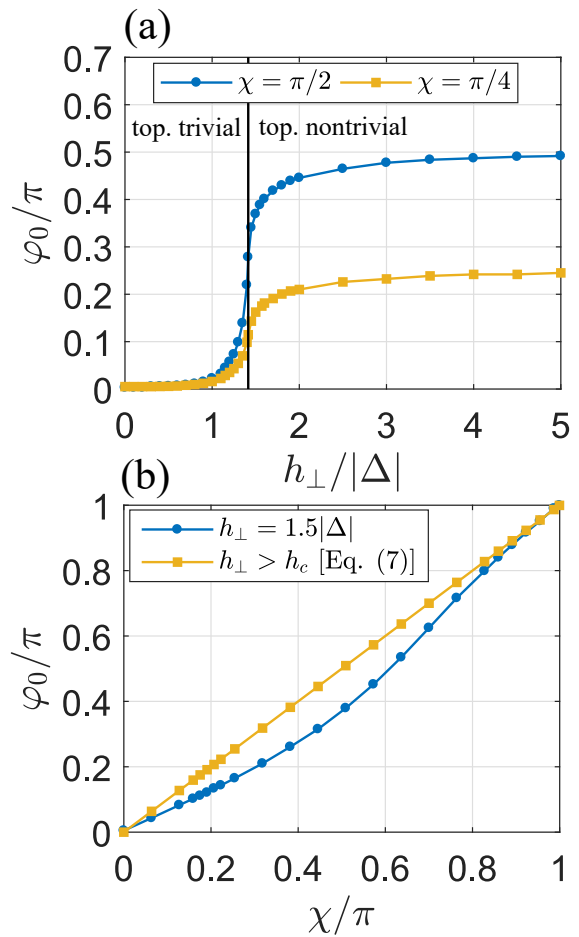}
\caption{\label{Fig:perpendicular_phase_offset} Typical $\varphi_0(h_{\perp})$ dependencies (a) and $\varphi_0(\chi)$ plots (b) for the model profile of the Zeeman field~(\ref{Zeeman_profile_perpendicular}). We take $\mu = |\Delta|$ and $m\alpha^2 = 0.2|\Delta|$ for both nanowires. The system undergoes the topological phase transition at $h_{\perp} = \sqrt{2}|\Delta|$ [denoted by a black solid line in panel (a)].}
\end{figure} 

\begin{figure}[htpb]
\centering
\includegraphics[scale = 0.9]{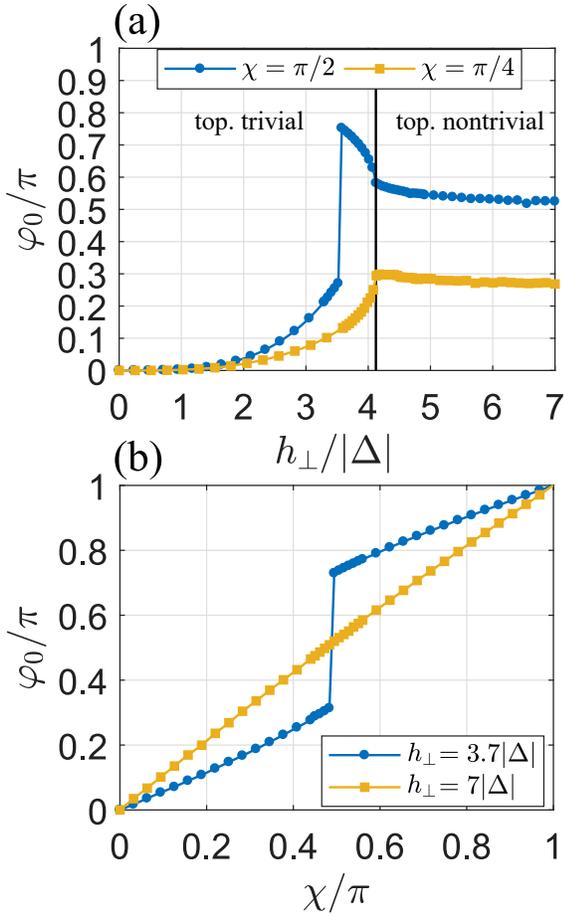}
\caption{\label{Fig:perpendicular_phase_offset_mu4} Typical $\varphi_0(h_{\perp})$ dependencies (a) and $\varphi_0(\chi)$ plots (b) for the model profile of the Zeeman field~(\ref{Zeeman_profile_perpendicular}). We take $\mu = 4|\Delta|$ and $m\alpha^2 = 0.2|\Delta|$ for both nanowires. The system undergoes the topological phase transition at $h_{\perp} = \sqrt{17}|\Delta|$ [denoted by a black solid line in panel (a)].}
\end{figure} 

Typical $\varphi_0(h_{\perp})$ dependencies within the full range of Zeeman fields covering both the topologically trivial and nontrivial regions of the phase diagram are shown in Figs.~\ref{Fig:perpendicular_phase_offset}(a) and~\ref{Fig:perpendicular_phase_offset_mu4}(a). Note that we take $\mu = |\Delta|$ and $\mu = 4|\Delta|$ to produce the plots in Figs.~\ref{Fig:perpendicular_phase_offset}(a) and~\ref{Fig:perpendicular_phase_offset_mu4}(a), respectively. Our results indicate that the anomalous phase shift vanishes for $h_{\perp}\lesssim |\Delta|$ while $\varphi_0 \approx \chi$ for $h_{\perp} \gg h_c$.  One can notice that in between these limiting cases $\varphi_0(h_{\perp})$ curves in Figs.~\ref{Fig:perpendicular_phase_offset}(a) and~\ref{Fig:perpendicular_phase_offset_mu4}(a) reveal qualitatively different behavior. Indeed, the plots in Fig.~\ref{Fig:perpendicular_phase_offset}(a) reveal a smooth crossover from the conventional to anomalous Josephson effect and the field of the topological transition corresponds to the maximum slope on $\varphi_0(h_{\perp})$ curves. The increase in the chemical potential $\mu$ results in a more sharp peculiarities at the field of the topological transition which now reveals itself through the appearance of the kink on $\varphi_0(h_{\perp})$ curves in Fig.~\ref{Fig:perpendicular_phase_offset_mu4}(a). Moreover, the results presented in Fig.~\ref{Fig:perpendicular_phase_offset_mu4}(a) for $\chi = \pi/2$ show the jump of the anomalous phase in the topologically trivial regime for $h_{\perp}\approx 3.5|\Delta|$. Such behavior stems from the presence of two local minima of the total energy of the contact vs. the superconducting phase difference. Thus, the jump of the anomalous phase appears exactly due to the competition of these two minima as one changes the Zeeman field. Typical dependencies of the anomalous phase shift on the geometrical offset angle are shown in Figs.~\ref{Fig:perpendicular_phase_offset}(b) and~\ref{Fig:perpendicular_phase_offset_mu4}(b). One can see that, quite opposite to the case of large Zeeman fields, typical $\varphi_0(\chi)$ dependencies are nonlinear within the crossover region on $\varphi_0(h_{\perp})$ curves. We emphasize that the above-described behavior of the anomalous phase shift is in qualitative agreement with the results of the effective BdG model~(\ref{simplified_equations}). 

\begin{figure}[htpb]
\centering
\includegraphics[scale = 1]{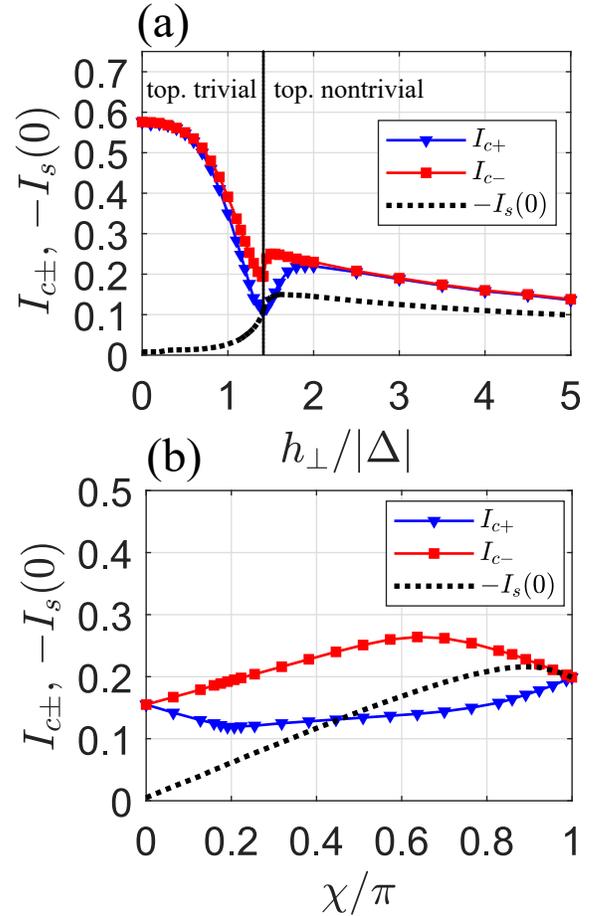}
\caption{\label{Fig:perpendicular_current} Dependencies of the critical currents $I_{c\pm}$ and the anomalous current on the Zeeman splitting (a) and on the geometrical offset angle (b) for the model profile of the Zeeman field given by Eq.~(\ref{Zeeman_profile_perpendicular}). We take $\mu = |\Delta|$ and $m\alpha^2 = 0.2|\Delta|$ for both nanowires, $\chi = \pi/2$ for (a) and $h_{\perp} = 1.5|\Delta|$ for (b). The system undergoes the topological phase transition at $h_{\perp} = \sqrt{2}|\Delta|$ [denoted by a black solid line in panel (a)].}
\end{figure}

In the end of this subsection we study the dependencies of key transport characteristics of the Josephson effect. Typical dependencies of the critical currents $I_{c\pm}$ and the anomalous Josephson current $I_s(\varphi = 0)$ on the Zeeman splitting and the geometrical offset angle are shown in Fig.~\ref{Fig:perpendicular_current}. One can see from Fig.~\ref{Fig:perpendicular_current}(a) that for $h_{\perp}\lesssim |\Delta|$ the system does not exhibit the diode effect $I_{c+}\approx I_{c-}$ and the anomalous current is fully suppressed. Increasing the Zeeman field from zero towards the topological transition leads to the decrease in the critical currents and to the increase in the absolute value of the anomalous current. The results shown in Fig.~\ref{Fig:perpendicular_current}(a) indicate that within the crossover region on $\varphi_0(h_{\perp})$ curves the magnitude of the critical current depends on the direction of the applied current $I_{c+}\neq I_{c-}$, thus, the system exhibits the diode effect. Moreover, the plots in Fig.~\ref{Fig:perpendicular_current}(a) reveal the jump of the derivatives $\partial I_{c\pm}/\partial h_{\perp}$ at the topological phase transition. Further increase in the spin splitting field in the topological phase reveals a
reentrant behavior of the critical currents. In accordance with the results obtained in Ref.~\cite{San-Hose2013}, we find that the above-mentioned reentrant behavior becomes less pronounced as one approaches the limit $\chi\to 0$. It follows both from Eq.~(\ref{supercurrent_analytical}) and our numerical simulations that in the limit of large Zeeman fields the critical current does not depend on the geometrical offset angle and the direction of the applied current $I_{c+}= I_{c-} = \mathcal{I}_-$. Comparing Eq.~(\ref{supercurrent_analytical}) for $h_{\perp}\gg h_c$ with $I_{c\pm}(\chi)$ dependencies calculated near the topological transition [see Fig.~\ref{Fig:perpendicular_current}(b)], one can see that the crossover region on $\varphi_0(h_{\perp})$ curves also reveals itself in the modification of angular dependencies of the critical currents. The above analysis of the behavior of the critical currents suggests that the singularity in the derivative of the critical current can be used to identify the topological transitions in proximized nanowire networks.

\subsection{Textured spin splitting field}

\begin{figure}[htpb]
\centering
\includegraphics[scale = 0.9]{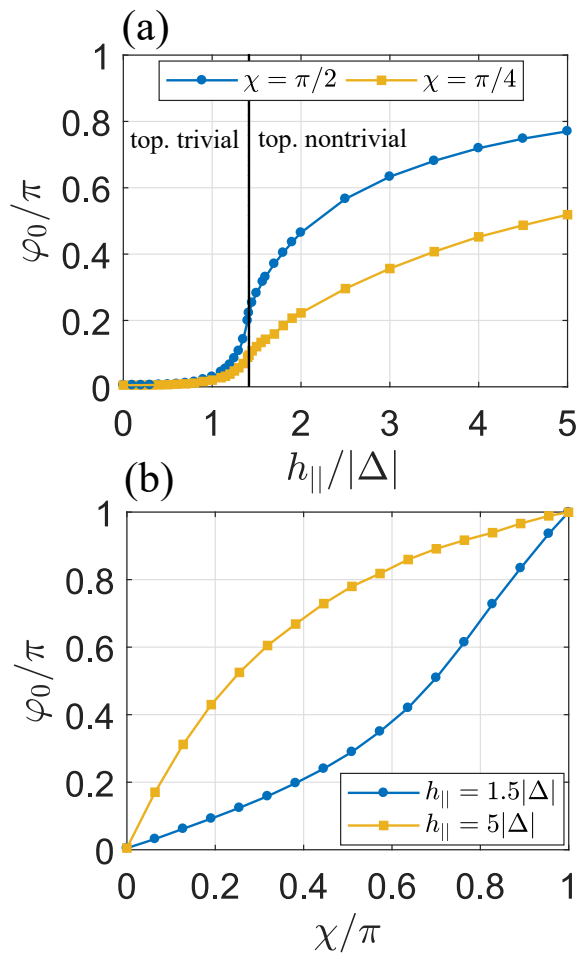}
\caption{\label{Fig:inplane_phase_offset} Typical $\varphi_0(h_{||})$ dependencies (a) and $\varphi_0(\chi)$ plots (b) for the model profile of the spin splitting field given by Eq.~(\ref{Zeeman_profile_inplane}). We take $\mu = |\Delta|$ and $m\alpha^2 = 0.2|\Delta|$ for both nanowires. The system undergoes the topological phase transition at $h_{||} = \sqrt{2}|\Delta|$ [denoted by a black solid line in panel (a)].}
\end{figure}

\begin{figure}[htpb]
\centering
\includegraphics[scale = 0.9]{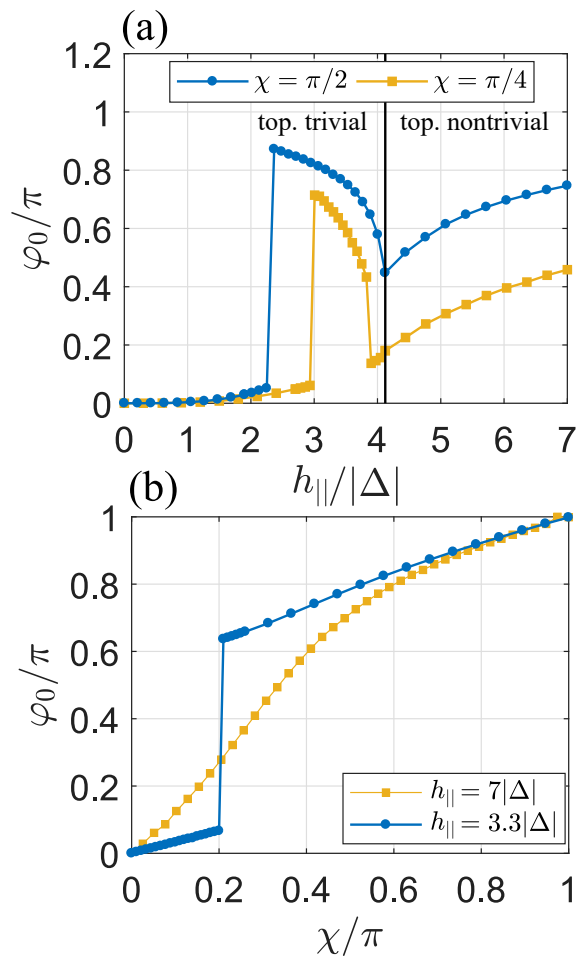}
\caption{\label{Fig:inplane_phase_offset_mu4} Typical $\varphi_0(h_{||})$ dependencies (a) and $\varphi_0(\chi)$ plots (b) for the model profile of the spin splitting field given by Eq.~(\ref{Zeeman_profile_inplane}). We take $\mu = 4|\Delta|$ and $m\alpha^2 = 0.2|\Delta|$ for both nanowires. The system undergoes the topological phase transition at $h_{||} = \sqrt{17}|\Delta|$ [denoted by a black solid line in panel (a)].}
\end{figure}

Here we consider the model profile of the spin splitting field given by Eq.~(\ref{Zeeman_profile_inplane}). Typical $\varphi_0(h_{||})$ dependencies are shown in Figs.~\ref{Fig:inplane_phase_offset}(a) and~\ref{Fig:inplane_phase_offset_mu4}(a). We use $\mu = |\Delta|$ and $\mu = 4|\Delta|$ for the plots in Figs.~\ref{Fig:inplane_phase_offset}(a) and~\ref{Fig:inplane_phase_offset_mu4}(a), respectively. One can clearly see that the crossover region on $\varphi_0(h_{||})$ curves extends far into the topologically nontrivial phase and the superconducting phase offset significantly deviates from the geometrical offset angle even for large spin splitting fields $h_{||}\gg h_c$. Our numerical simulations reveal that the contribution of the continuum states to supercurrent is comparable with the one from the subgap states even for large spin splitting fields. Thus, the analysis of the subgap states is not enough for the description of the junction characteristics. The results in Figs.~\ref{Fig:inplane_phase_offset}(a) and~\ref{Fig:inplane_phase_offset_mu4}(a) show a number of peculiarities in the behavior of the anomalous phase. Note that $\varphi_0(h_{||})$ curves in Fig.~\ref{Fig:inplane_phase_offset}(a) (for $\mu = |\Delta|$) reveal a smooth crossover from the conventional to the anomalous Josephson effect while the results in Fig.~\ref{Fig:inplane_phase_offset_mu4}(a) (for $\mu = 4|\Delta|$) indicate the jumps of the anomalous phase in the topologically trivial region. In particular, the plots in Fig.~\ref{Fig:inplane_phase_offset_mu4}(a) for $\chi = \pi/2$ indicate a jump of the anomalous phase at $h_{||}\approx 2.2|\Delta|$ and a dip at the topological transition. The appearance of the jumps arises from the interplay of two competing local minima of the total energy vs. the superconducting phase difference. The results presented in Fig.~\ref{Fig:inplane_phase_offset_mu4}(a) for $\chi = \pi/4$ reveal even two jumps of the anomalous phase at $h_{||}\approx 3|\Delta|$ and $h_{||}\approx 4|\Delta|$ in the topologically trivial phase. 
Typical dependencies of the anomalous phase shift on the geometrical offset angle for the spin splitting profile~(\ref{Zeeman_profile_inplane}) are shown in Figs.~\ref{Fig:inplane_phase_offset}(b) and~\ref{Fig:inplane_phase_offset_mu4}(b). One can see from Fig.~\ref{Fig:inplane_phase_offset}(b) and~\ref{Fig:inplane_phase_offset_mu4}(b) that $\varphi_0(\chi)$ dependencies are nonlinear both in the topologically trivial and nontrivial regimes.

\begin{figure}[htpb]
\centering
\includegraphics[scale = 1]{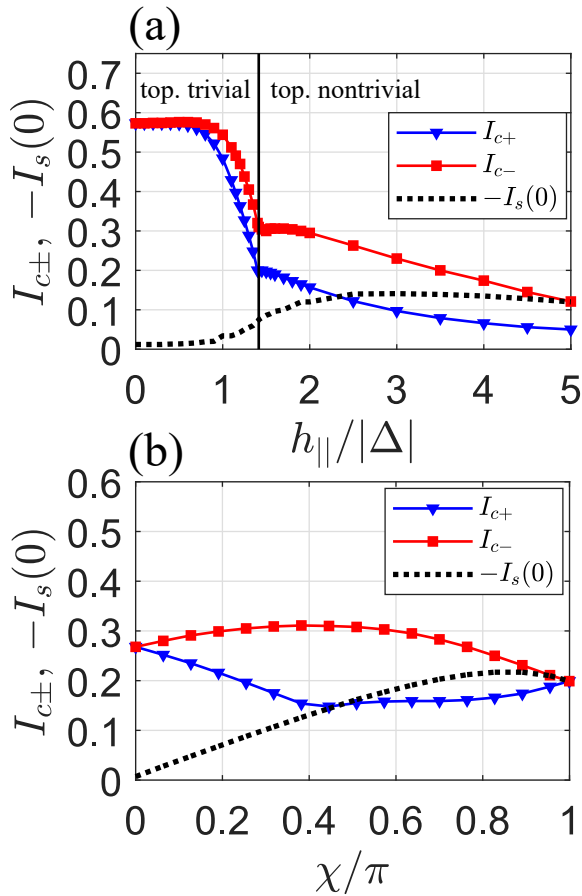}
\caption{\label{Fig:inplane_current} Dependencies of the critical currents $I_{c\pm}$ and the anomalous current on the spin splitting field (a) and on the geometrical offset angle (b) for the model profile of the spin splitting field given by Eq.~(\ref{Zeeman_profile_inplane}). We take $\mu = |\Delta|$ and $m\alpha^2 = 0.2|\Delta|$ for both nanowires, $\chi = \pi/4$ for (a) and $h_{||} = 1.5|\Delta|$ for (b). The system undergoes the topological phase transition at $h_{||} = \sqrt{2}|\Delta|$ [denoted by a black solid line in panel (a)].}
\end{figure}

Typical dependencies of the critical currents $I_{c\pm}$ and the anomalous Josephson current $I_s(\varphi = 0)$ on the spin splitting field and the geometrical offset angle are shown in Fig.~\ref{Fig:inplane_current}. These results demonstrate that all previously mentioned features of the critical currents such as the jump of the derivative $\partial I_{c\pm}/\partial h_{||}$ at the topological transition and the superconducting diode effect $I_{c+}\neq I_{c-}$ are also observed in the case of the textured spin splitting field~(\ref{Zeeman_profile_inplane}). 

Finally, we argue that the above-described features of the Josephson transport originate due to the presence of two contributions to the total supercurrent from the subgap states and the continuum levels. Indeed, a necessary condition for the appearance of the geometry-induced effects under consideration is that these two contributions should be of different magnitude and have a different behavior with respect to the superconducting phase difference.

\section{Discussion}\label{discussion}

Experimentally, the characteristics of the anomalous Josephson effect studied in our work can be probed in various SQUID setups (see, e.g., Fig.~2 and the corresponding discussion in Ref.~\cite{Kutlin2020}). However, one can expect that the observation of the anomalous Josephson effect in systems based on InAs/Al nanowires should be quite challenging. First, typical magnetic fields required to drive such systems into the topologically nontrivial phase $H\sim 1$~T usually exceed perpendicular critical magnetic fields for the destruction of superconductivity in Al shell $H_{c\perp} \sim 0.1$~T~\cite{Chang2015,Higginbotham2015}. The second problem is that the observation of the anomalous Josephson effect in curved nanowire junctions with a textured profile of the spin splitting field shown in Fig.~\ref{Fig:model_system}(b) requires a precise alignment of external magnetic fields at nanometer scales. We anticipate that both these problems should be naturally resolved in a novel type of Majorana devices with epitaxial layer of superconducting Al and the ferromagnetic insulator EuS grown on InAs nanowires~\cite{Vaitikenas2020}. Existing experimental data reveal that EuS becomes magnetized along the wire axis with typical switching (coercive) field of  $\pm 11$~mT and an inferred remanent Zeeman field of $\sim 1.3$~T. The fact that EuS magnetizes along the wire axis makes these devices quite promising for the realization of topological nanowire networks since their use eliminates the need for precise alignment of external magnetic fields. Thus, we argue that ferromagnetic hybrid nanowires are a natural platform to probe the peculiarities of the Josephson effect studied in our work. Moreover, the analysis of various characteristics of the Josephson effect (the anomalous phase offset shift, critical current) on the model parameters carried out in our work can be used to estimate the parameters of such devices (the strength of spin-orbit coupling as well as the induced Zeeman field in the semiconducting core) crucial for the relalization of the topological superconductivity in hybrid nanowires.

Let us note in conclusion that the geomtry-induced transport effects considered in our work can be simulated in straight nanowire junctions with an inhomogeneous spin-orbit coupling. Indeed, for the spin-orbit term in our model Hamiltonian~(\ref{BdG_matrix}) the spin operator $\hat{\boldsymbol{\sigma}}$ is projected onto the axis $\mathbf{n} \times \boldsymbol{\tau}(s)$ which changes its direction along the curved wire. Here $\mathbf{n} = [0,0,1]$ is the Rashba unit vector and $\boldsymbol{\tau} = [\cos\chi(s),\sin\chi(s),0]$ is the unit vector along the wire axis. One can see that choosing the appropriate profile of the Rashba unit vector $\mathbf{n}(s)$ along a straight wire allows one to reduce the problem for a curved wire to the one for a straight wire with an inhomogeneous spin-orbit coupling. However, the experimental realization of such interaction requires a precise control of the direction of built-in electric fields in the sample and is quite challenging (see Refs.~\cite{Ojanen2013,Klinovaja2015} and references therein).

\section{Summary}\label{summary}
To sum up, we have uncovered and explained the crossover between the conventional and the anomalous Josephson effect in curved nanowire junctions as both proximitized nanowire parts undergo a transition to the topologically nontrivial state. We have shown that both the subgap and continuum quasiparticle states are responsible for such unusual behavior of the anomalous Josephson phase. We have investigated the manifestations of the above crossover on the dependencies of the critical current on the spin splitting and geometrical offset angle. In particluar, we have demonstrated that the above crossover reveals itself in the superconducting diode effect: the magnitude of the critical current depends on the direction of the applied current. Our results suggest a new type of a tunable phase battery which can be experimentally implemented in systems based on hybrid ferromagnetic nanowires.
The resulting phase battery can be used as a probe of topological transitions in Majorana networks and can become a useful element of various quantum computation devices.

\acknowledgments
We thank A.A.~Bespalov, S.V.~Mironov, D.Yu.~Vodolazov, V.L.~Vadimov, M.A.~Silaev, I.M.~Khaymovich and P.I.~Arseyev for valuable comments. This work was supported in part by the Russian Foundation for Basic Research under Grant No. 19-31-51019 and the Russian State Contract No. 0035-2019-0021. The work involving numerical calculations for the case of a textured spin splitting field was supported by the Russian Science Foundation (Grant No. 20-12-00053).

\appendix

\section{Details of numerical simulations}\label{details_numerical_simulations}

Here we provide the details of numerical simulations. 
Our starting point is the BdG equations~(\ref{BdG_general}) and the geometry of the problem is shown schematically in Fig.~\ref{Fig:model_system} (without loss of generality, we assume that the contact is located at $s = 0$).
For numerical calculations it is convenient to exploit a formal equivalence between the inhomogeneous Rashba spin-orbit interaction and the inhomogeneous spin splitting field~\cite{Braunecker2010,Kjaergaard2012}. Performing the rotation of the BdG matrix~(\ref{BdG_matrix}) in the spin space $\check{\mathcal{H}}(s) = \check{U}^{\dagger}(s)\check{H}_{\text{BdG}}(s)\check{U}(s)$ where
\begin{equation}\label{rotation_operator}
 \check{U}(s) = \exp\left[ik_{so}\int_0^sds' \ \hat{\sigma}_n(s')\right] \ ,
\end{equation}
we get the following eigenvalue problem:
\begin{subequations}\label{BdG_rotated}
\begin{align}
 \check{\mathcal{H}}(s)\tilde{\Psi}(s) = E\tilde{\Psi}(s) \ ,\\
 \check{\mathcal{H}}(s) = \left[\frac{p^2}{2m} - \tilde{\mu}(s) \right]\check{\tau}_z + \mathbf{h}(s)\left[\check{U}^{\dagger}(s)\hat{\boldsymbol{\sigma}}\check{U}(s)\right]\\
 \nonumber
 + |\Delta|\left[\tau_x\cos\varphi(s) - \tau_y\sin\varphi(s)\right] .
\end{align}
\end{subequations}
Here $k_{so} = m\alpha/\hbar$, $\tilde{\mu}(s) = \mu(s) + m\alpha^2/2$, and $\tilde{\Psi}(s) = \check{U}^{\dagger}(s)\Psi(s)$. Substituting a step-wise profile for the geometrical phase $\chi(s) = \chi\Theta(s)$ into Eq.~(\ref{rotation_operator}), we get
\begin{eqnarray}\label{rotation_operator_step_wise_phase}
 \check{U}(s) = \left[\cos\left(k_{so}s\right) + i\sin\left(k_{so}s\right)\hat{\sigma}_n(s)\right] \ .
\end{eqnarray}
Eqs.~(\ref{BdG_rotated}) imply that the the rotated wave function and its derivative must be continuous at the contact which allows one to construct a simple finite-difference approximation for the eigenvalue problem~(\ref{BdG_rotated}). On the other hand, the boundary conditions for the initial problem~(\ref{BdG_general}) include the continuity of the quasiparticle wave function as well as the jump of its derivative at the contact 
\begin{subequations}\label{boundary_conditions_initial}
\begin{align}
\frac{d\Psi}{ds}\biggl|_{+0} - \frac{d\Psi}{ds}\biggl|_{-0} = \check{M}\Psi(0) \ , \\
 \check{M} = ik_{so}\left[\hat{\sigma}_x \sin\chi - \hat{\sigma}_y\left(\cos\chi - 1\right)\right] \ .
\end{align}
\end{subequations}
The above equations clearly show that a step-wise profile of the geometrical phase introduces an additional spin-dependent barrier for quasiparticles at the contact and the barrier strength is controlled both by the strength of spin-orbit interaction and the geometrical phase difference.

Finally, we describe the finite-difference scheme which we use for calculations of the quasiparticle spectra. Introducing the spatial grid $s_j = aj$, where $a$ is the discretization step size and $j$ is an integer, the finite-difference approximation of the eigenvalue problem~(\ref{BdG_rotated}) is as follows
\begin{subequations}
\begin{align}
 \sum_{l' = 1}^N\check{\mathcal{H}}_{ll'}\tilde{\Psi}(s_{l'}) = E\tilde{\Psi}(s_l) \ ,\\
 \label{BdG_matrix_fd}
 \check{\mathcal{H}}_{ll'} = \bigl\{\left[2t - \tilde{\mu}(s_l)\right]\check{\tau}_z + \mathbf{h}(s_l)\left[\check{U}^{\dagger}(s_l)\hat{\boldsymbol{\sigma}}\check{U}(s_l)\right] \\
 \nonumber
 + |\Delta|\left[\check{\tau}_x\cos\varphi(s_l) - \check{\tau}_y\sin\varphi(s_l)\right]\bigl\}\delta_{l,l'} \\
 \nonumber
 -t\check{\tau}_z(\delta_{l,l'-1} + \delta_{l,l'+1}) \ .
\end{align}
\end{subequations}
Here $N$ is the total number of grid points and $t = \hbar^2/2ma^2$. In our numerical simulations we $\ell \sqrt{m|\Delta|/\hbar} = 44$, $N = 1000$ and compute $400$ positive eigenvalues of the BdG matrix~(\ref{BdG_matrix_fd}). Here $\ell$ is the length of each nanowire. In order to exclude the influence of finite-size effects on our results, we perform the robustness check by increasing the wire length and the number of grid points. We have checked that a further doubling of $\ell$ with the same density of points yields only few percent deviations for the critical supercurrent and insignificant deviations for the anomalous phase shift.

\section{BdG equations in the helical basis and the effective low-energy model~(\ref{BdG_subbands_simplified})}\label{appendix_B}

To derive the effective model~(\ref{BdG_subbands_simplified}) we follow Ref.~\cite{Alicea2} and consider, first, an auxillary translation invariant problem for a straight nanowire in the momentum representation 
\begin{subequations}\label{BdG_translation_invariant}
    \begin{align}
        \check{H}(k)\Psi(k) = E\Psi(k) \ ,\\
        \check{H}(k) = \xi_k\check{\tau}_z - h_{\perp}\hat{\sigma}_z \\
        \nonumber
        + |\Delta|\left[\check{\tau}_x\cos\varphi - \check{\tau}_y\sin\varphi\right] - \lambda_k\hat{\sigma}_n(\chi)\check{\tau}_z \ .
    \end{align}
\end{subequations}  
Here $\hbar k$ is the momentum along the wire, $\xi_k = \hbar^2k^2/2m - \mu$, and $\lambda_k = \alpha \hbar k$. The above problem can be rewritten in the basis of the normal-state Hamiltonian
\begin{equation}
 \hat{H}_0 = \left[\xi_k - h_{\perp}\hat{\sigma}_z - \lambda_k\hat{\sigma}_n(\chi)\right] \ .
\end{equation}
As a result, one gets the following equations 
\begin{subequations}\label{BdG_helical_basis}
 \begin{align}
 \check{\mathcal{H}}_{hs}(k)\Psi_{hs}(k) = E \Psi_{hs}(k) \ ,\\
 \check{\mathcal{H}}_{hs}(k) = \begin{bmatrix}\varepsilon_+(k)&\Delta_{p+}(k)&0&\Delta_{s}(k)\\ \Delta_{p+}^{\dagger}(k)&-\varepsilon_+(k)&\Delta_{s}^{\dagger}(k)&0\\0&\Delta_{s}(k)&\varepsilon_-(k)&\Delta_{p-}(k)\\ \Delta_{s}^{\dagger}(k)&0&\Delta_{p-}^{\dagger}(k)&-\varepsilon_-(k)\end{bmatrix} \ ,
 \end{align}
\end{subequations}
where $\varepsilon_{\pm}(k) = \xi_k \pm \sqrt{h_{\perp}^2 + \lambda_k^2}$ is the energy spectrum of the normal-state helical subbands and $\Psi_{hs} = [u_+,v_+,u_-,v_-]^{{\rm T}}$ is the quasiparticle wave function in the helical basis. One can clearly see from Eqs.~(\ref{BdG_helical_basis}) that the eigenvalue problem~(\ref{BdG_translation_invariant}) for the spinful proximitized nanowire is reduced to two one-dimensional superconductors with the $p$-wave intraband pairing
\begin{subequations}
 \begin{align}
  \Delta_{p+}(k) = |\Delta|e^{i(\varphi + \chi)}\frac{i\lambda_k}{\sqrt{h_{\perp}^2 + \lambda_k^2}} \ ,\\
  \Delta_{p-}(k) = |\Delta|e^{i(\varphi - \chi)}\frac{i\lambda_k}{\sqrt{h_{\perp}^2 + \lambda_k^2}} 
 \end{align}
\end{subequations}
coupled through the interband $s$-wave gap function
\begin{equation}
 \Delta_{s}(k) = |\Delta|e^{i\varphi}\frac{h_{\perp}}{\sqrt{h_{\perp}^2 + \lambda_k^2}} \ .
\end{equation}
Within the limit $\mu \gg |\Delta|,m\alpha^2$ and $h_{\perp}\gtrsim |\Delta|$ one can neglect the interband $s$-wave pairing~\cite{Alicea2,San-Hose2013,Kutlin2020}. The resulting block diagonal BdG matrix has the following form:
\begin{subequations}
 \begin{align}
  \check{\mathcal{H}}_{hs}(k) = \begin{bmatrix}\check{H}_+(k)&0\\0&\check{H}_-(k)\end{bmatrix} \ ,\\
  \label{effective_Hamiltonian}
  \check{H}_{\eta}(k) = \begin{bmatrix}\xi_k + \eta h_{\perp}&\cfrac{i\lambda_k|\Delta|}{h_{\perp}}e^{i(\varphi + \eta\chi)}\\-\cfrac{i\lambda_k|\Delta|}{h_{\perp}}e^{-i(\varphi + \eta\chi)}&-\xi_k-\eta h_{\perp}\end{bmatrix} \ .
 \end{align}
\end{subequations}
Here $\eta = \pm$ is the subband index. Generalization of the above equation to the inhomogeneous case can be obtained either following the derivation, e.g., in Ref.~\cite{Kutlin2020} or just replacing $k \to -i\partial_s$ in the above matrices and introducing the appropriate anticommutators in the gap operators to allow spatial variations of the superconducting phase difference $\varphi(s)$ and the geometrical offset angle $\chi(s)$. As a result, we get the effective BdG model~(\ref{BdG_subbands_simplified}).

\
\\
\

\section{Derivation of Eq.~(\ref{subgap_spectrum}) in the main text}\label{equations_derivation}
Here we provide the derivation of Eq.~(\ref{subgap_spectrum}) in the main text~\cite{Spanslatt2018, Kutlin2020}. Our starting point is the BdG equations~(\ref{simplified_equations})
\begin{equation}\label{BdG_subband_equations_appendix}
 \begin{bmatrix}\xi(s) + \eta h_{\perp} & \Delta_{p\eta}(s)\\ \Delta_{p\eta}^{\dagger}(s)&-\xi(s) - \eta h_{\perp}\end{bmatrix}\begin{bmatrix}u_{\eta}(s)\\ v_{\eta}(s)\end{bmatrix} = E\begin{bmatrix}u_{\eta}(s)\\ v_{\eta}(s)\end{bmatrix} 
\end{equation}
and the $p$-wave gap operator $\Delta_{p\eta}(s)$ is given by Eq.~(\ref{pairing_terms_simplified}). Substituting step-wise profiles for both superconducting and geometrical phases into Eq.~(\ref{BdG_subband_equations_appendix}), one can easily get the solutions in the quasiclassical approximation
\begin{subequations}\label{quasiclassical_subband_ansatz}
 \begin{align}
  \begin{bmatrix}u_{\eta}(s)\\v_{\eta}(s)\end{bmatrix}_{s<0} = 
  e^{\zeta_{\eta}s}\biggl\{a_{\eta +}e^{ik_{F\eta}s}\begin{bmatrix}1\\ -ie^{i\gamma_{\eta}}\end{bmatrix}\\
  \nonumber
  + a_{\eta -}e^{-ik_{F\eta}s}\begin{bmatrix}1\\ ie^{-i\gamma_{\eta}}\end{bmatrix}\biggl\} \ ,\\
  \begin{bmatrix}u_{\eta}(s)\\v_{\eta}(s)\end{bmatrix}_{s>0} = e^{-\zeta_{\eta}s}\biggl\{b_{\eta +}e^{ik_{F\eta}s}\begin{bmatrix}1\\ -ie^{-i\gamma_{\eta}-i(\varphi + \eta\chi)}\end{bmatrix} \\
  \nonumber
  + b_{\eta-}e^{-ik_{F\eta}s}\begin{bmatrix}1\\ ie^{i\gamma_{\eta}-i(\varphi + \eta\chi)}\end{bmatrix}\biggl\} \ .
 \end{align}
\end{subequations}
Here $\hbar k_{F\eta} = {\rm Re}[\sqrt{2m(\mu-\eta h_{\perp})}]$ is the subband Fermi momentum, $\zeta_{\eta} = \sqrt{|\Delta_{\eta}|^2-E^2}/\hbar v_{F\eta}$, $|\Delta_{\eta}| = \alpha p_{F\eta}|\Delta|/h_{\perp}$ is the superconducting gap in the spectrum of the subband $\eta$, $v_{F\eta}$ is the Fermi velocity, $\gamma_{\eta} = \arccos(E/|\Delta_{\eta}|)$, and $|E|<|\Delta_{\eta}|$. Matching the quasiparticle wave functions~(\ref{quasiclassical_subband_ansatz}) in the case of transparent junction, we get the spectral equation
\begin{equation}
 \sin\left[\gamma_{\eta} + (\varphi + \eta\chi)/2\right] = 0 \ .
\end{equation}
The solutions of the above equation are given by Eq.~(\ref{subgap_spectrum}). Note also that the condition for the validity of the quasiclassical approximation $\hbar^2k_{F\eta}^2/2m\gg |\Delta|$ breaks down for $\eta = +$ subband in the vicinity of the topological transition since $k_{F+}\to 0$ as $\mu \to h_{\perp}$.


\end{document}